\documentclass[conference]{IEEEtran}
\IEEEoverridecommandlockouts

\usepackage{etoolbox}
\usepackage{multicol}
\usepackage{multirow}
\makeatletter
\patchcmd{\@makecaption}
  {\scshape}
  {}
  {}
  {}
\makeatother

\usepackage{cite}
\usepackage{amsmath,amssymb,amsfonts}
\usepackage{mathtools}
\newcommand{\norm}[1]{\left\lVert#1\right\rVert}

\usepackage{algorithmic}
\usepackage{graphicx}
\usepackage{textcomp}

\usepackage{array, booktabs, makecell}
\usepackage{siunitx, mhchem}

\usepackage{xcolor}
\usepackage[numbers]{natbib}
\def\BibTeX{{\rm B\kern-.05em{\sc i\kern-.025em b}\kern-.08em
    T\kern-.1667em\lower.7ex\hbox{E}\kern-.125emX}}
\begin{document}



\newcommand{\papername}{NAX }
\newcommand{\papernamewospace}{NAX}




\title{\papernamewospace: Co-Designing \underline{N}eural Network and Hardware \underline{A}rchitecture for Memristive \underline{X}bar based Computing Systems}





\author{\IEEEauthorblockN{Shubham Negi}
\IEEEauthorblockA{ 
\textit{Purdue University}\\
snegi@purdue.edu}
\and
\IEEEauthorblockN{Indranil Chakraborty}
\IEEEauthorblockA{
\textit{Purdue University}\\
ichakra@purdue.edu}
\and
\IEEEauthorblockN{Aayush Ankit\textsuperscript{*}}
\IEEEauthorblockA{
\textit{Microsoft Corporation}\thanks{ \textsuperscript{*}Work done while at Purdue University}\\
aayushankit@microsoft.com}
\and
\IEEEauthorblockN{Kaushik Roy}
\IEEEauthorblockA{
\textit{Purdue University}\\
kaushik@purdue.edu}
}

\maketitle
\begin{abstract}

In-Memory Computing (IMC) hardware using Memristive Crossbar Arrays (MCAs) are gaining popularity to accelerate Deep Neural Networks (DNNs) since it alleviates the "memory wall" problem associated with von-Neumann architecture. The hardware efficiency (energy, latency and area) as well as application accuracy (considering device and circuit non-idealities) of DNNs mapped to such hardware are co-dependent on network parameters, such as kernel size, depth etc. and hardware architecture parameters such as crossbar size. 
However, co-optimization of both network and hardware parameters presents a challenging search space comprising of different kernel sizes mapped to varying crossbar sizes.

To that effect, we propose \textit{\papernamewospace} -- an efficient neural architecture search engine that co-designs neural network and IMC based hardware architecture. \papername explores the aforementioned search space to determine kernel and corresponding crossbar sizes for each DNN layer to achieve optimal tradeoffs between hardware efficiency and application accuracy. Our results from \papername show that the networks have heterogeneous crossbar sizes across different network layers, and achieves optimal hardware efficiency and accuracy considering the non-idealities in crossbars. On CIFAR-10 and Tiny ImageNet, our models achieve 0.8\%, 0.2\% higher accuracy, and 17\%, 4\% lower \textit{EDAP} (energy-delay-area product) compared to a baseline ResNet-20 and ResNet-18 models, respectively.

\end{abstract}


\section{Introduction}
Deep Neural Networks (DNNs) are widely used in a variety of applications like object detection/recognition, video analytics, natural language processing etc. However, the increase of model size to achieve higher accuracy has led to a growing interest  
in designing domain-specific accelerators like Google TPU \cite{jouppi2017datacenter}, Microsoft BrainWave \cite{fowers2018configurable}, Intel Nervana NNP \cite{9154492} which accelerate the key compute primitive -- Matrix Vector Multiplication (MVM) \cite{verma2019memory} in the DNNs. One key aspect of these accelerators is to amortize the data-movement cost in von-Neumann architectures by bringing computation closer to memory. To that effect, researchers have explored Memristive Crossbar Array (MCA) \cite{ chakraborty2020resistive} based In-Memory Computing (IMC) hardware \cite{ankit2019puma, shafiee2016isaac, chi2016prime} which performs analog computations inside the memory array itself. Owing to the high on-chip density of MCAs \cite{hu2018memristor} and their ability to perform efficient in-situ MVM operations \cite{hu2016dot}, MCA based IMC hardware can achieve higher energy-efficiency compared to digital hardwares.



MCA based IMC hardware offer tremendous potential to accelerate machine learning (ML) workloads \cite{ankit2019puma, shafiee2016isaac}. However, there are significant challenges to overcome  \cite{ankit2020circuits}. In particular, MVM operations are performed in the analog domain and hence require analog-to-digital converters (ADCs) to communicate digital outputs across several MCAs. ADC's contribute majorly to the energy, latency and area of IMC hardware \cite{ankit2020circuits}. Since ADC precision depends on the crossbar size \cite{ankit2019puma}, the hardware efficiency (energy, latency and area) of such hardwares has a strong dependence on the crossbar size. In addition, MCA also suffers from computational errors originating from device and circuit non-idealities such as: parasitic resistance, non-linearity from access transistors and I-V characteristics of NVM device \cite{chakraborty2020geniex}. The degree of error in computation also depends on crossbar size \cite{chakraborty2020geniex}. In large-scale DNNs, these computation errors can accumulate and result in severe degradation in classification accuracy. Consequently, efficient mapping of DNN model to MCA based IMC hardware requires careful consideration of hardware design parameter such as crossbar size.


\begin{figure}[t]
\centering
\includegraphics[width=0.48\textwidth]{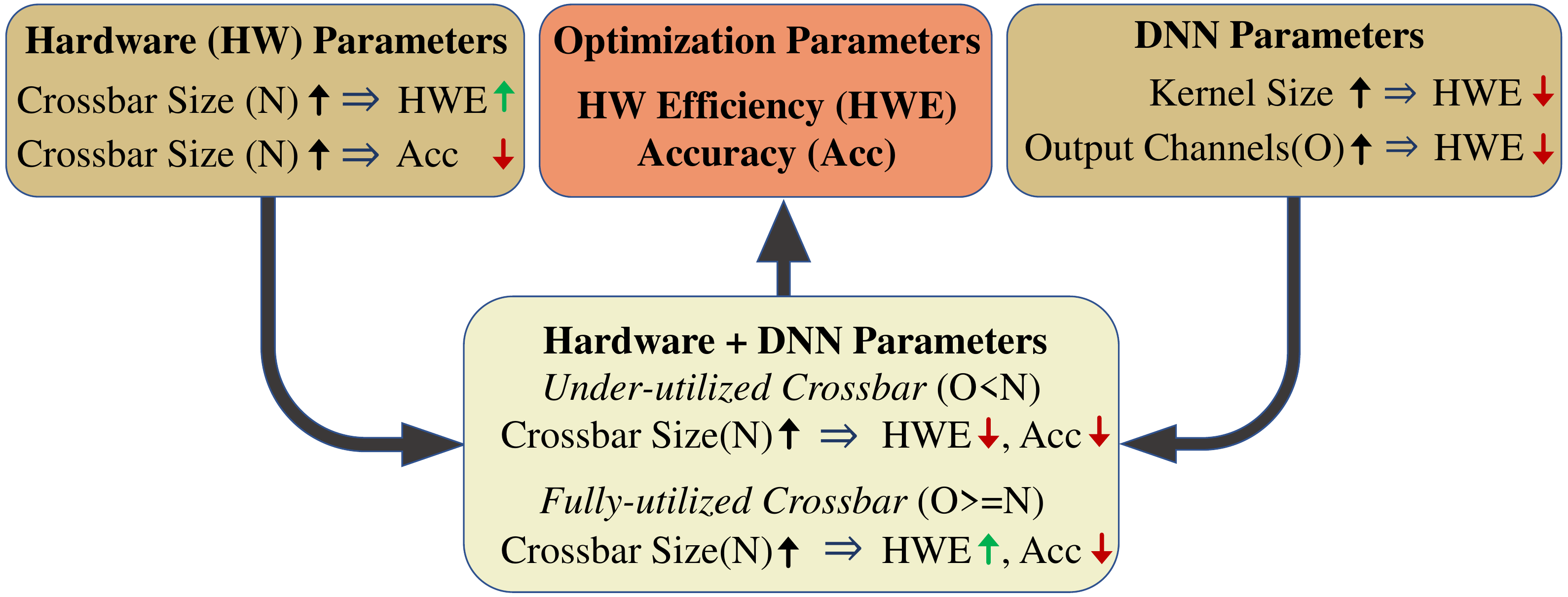}
\vspace*{-3mm}
\caption{Co-designing neural network and analog IMC hardware require exploration of a complex search space.}

\centering
\label{logical}
\end{figure}



\begin{figure*}[t]
\centering
\includegraphics[width=0.9\textwidth]{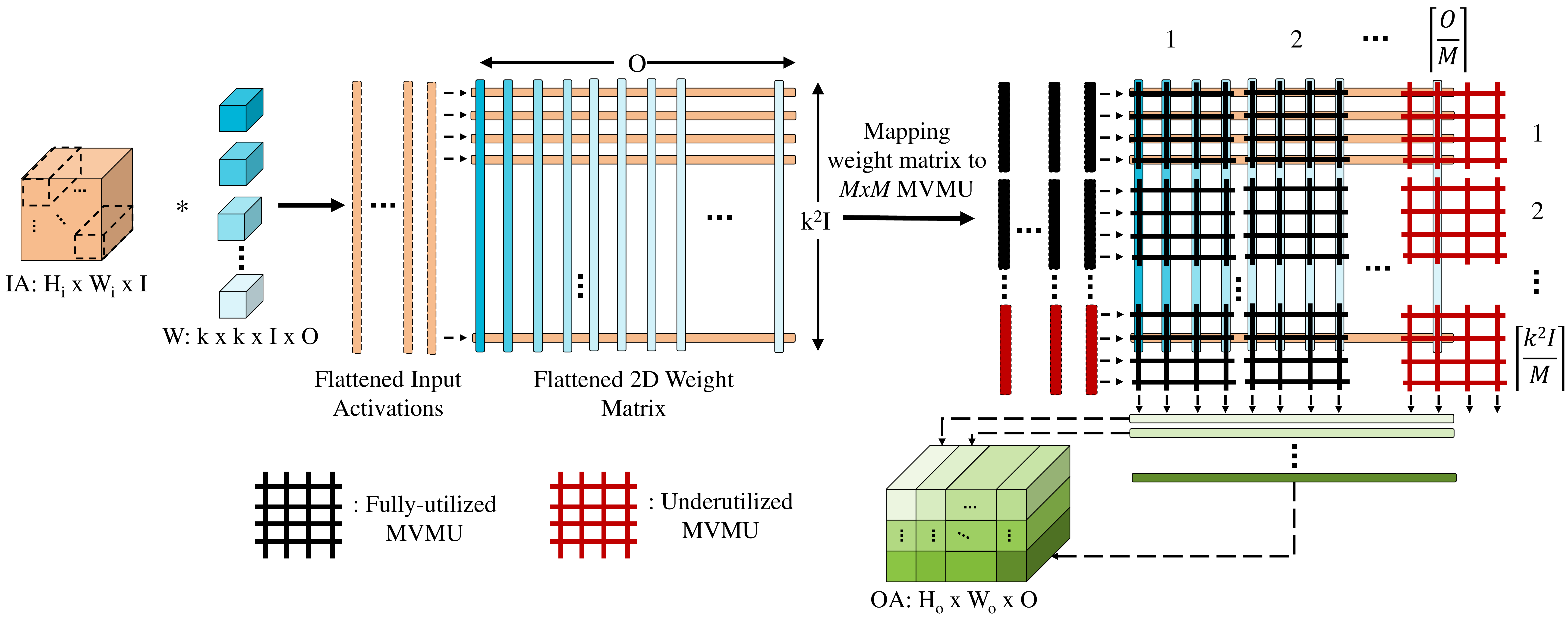}
\vspace*{-3mm}
\caption{Illustration of kernels of a DNN mapped to the MVMUs.  MVMU is shown with a single crossbar for simplicity.}

\centering
\label{underutilization}
\end{figure*}

The hardware efficiency and accuracy of DNN models primarily depends on its kernel and layer dimensions, when deployed on fixed hardware  configurations (CPUs, GPUs, mobile devices) \cite{zhou2021rethinking, cai2018proxylessnas, tan2019mnasnet}. However, in order to efficiently map DNNs to IMC hardware, we need to not only consider these DNN model parameters, but also the underlying hardware design parameter such as crossbar size. Intuitively, larger crossbars amortize peripheral (ADC) cost, which should lead to better hardware efficiency. However, for some layer operations, this may not be the case. For instance, as shown in Fig.~\ref{logical}, an operation (with O output channels) when mapped to a crossbar of size NxN (N$>$O), can lead to under-utilization of the crossbar. In such a case, a smaller crossbar of size MxM (N$>$M=O) which the operation fully utilizes can achieve better hardware efficiency for that particular operation. On the other hand, the accuracy of the DNN reduces with increasing crossbar size. This is because the effect of non-idealities is higher in larger crossbars \cite{chakraborty2020geniex}. Therefore, in order to maximize the hardware efficiency and accuracy of the DNN mapped on analog IMC hardware, there is a need to perform co-exploration of both DNN model parameters and hardware design parameters. However, such a co-design presents a challenging search space (as depicted in Fig.~\ref{logical}) consisting of both DNN and hardware design parameters.

Neural Architecture Search (NAS) has been widely used to explore over a large search space in order to design efficient neural networks for various deep learning tasks \cite{zoph2016neural, chen2019progressive, cai2018proxylessnas}. More recently, there has been an increasing interest in utilizing NAS frameworks to design neural networks catered towards hardware platforms \cite{abdelfattah2020best, yang2020co, choi2020dance} but such works have primarily focused on digital accelerators.
In this work, we present \papernamewospace, a NAS framework to design neural networks optimized for MCA based IMC hardware through co-exploration
of both DNN model parameters and the underlying hardware architecture parameters such as crossbar size. To the best of our knowledge, \papername is the first work that considers the crossbar size and all the significant non-idealities in crossbars to co-design DNN and IMC hardware. Our main contributions are as follows:



\begin{itemize}

    \item Analyze the unique co-dependence of the DNN model parameters such as kernel size, layer depth etc and hardware parameters such as crossbar size through their implications on hardware efficiency and accuracy of IMC hardware (section \ref{motivation}).
    
    \item Developed a NAS framework that explores the search space consisting of DNN model and hardware design parameters to co-design efficient neural network and analog IMC hardware (section \ref{section4}). 
    

    

    \item Perform an ablation study to demonstrate the importance of co-optimization of accuracy, energy and latency (area normalized) to design efficient neural networks for IMC hardware (section \ref{experiments}). 
    



\end{itemize}

\section{Related Work}

Past works have explored hardware-aware NAS \cite{cai2018proxylessnas, tan2019mnasnet} focused towards optimizing DNN model parameters for performance and accuracy on target platforms like mobile devices, GPUs etc.
More recently, several works \cite{abdelfattah2020best, yang2020co, choi2020dance, zhou2021rethinking, jiang2020device} have explored the co-design of DNNs and digital hardware accelerators using NAS-based frameworks.
In contrary, the search space for analog IMC hardware is unique and complex as DNN model parameters and the underlying hardware architecture affect both hardware efficiency and application accuracy.

With regard to analog IMC hardwares, past literature has been limited to optimization of DNN model parameters in isolation \cite{jiang2020device, parsa2019pabo}
without consideration of their co-dependence with the underlying hardware architecture in determination of the hardware efficiency and accuracy. 
Naturally, there remains a need to perform a scaled co-exploration of DNN model parameters as well as hardware parameters such as crossbar size to designing efficient DNNs for analog IMC hardwares. Our proposal \papername is a NAS-based framework to co-design DNNs as well as the underlying analog IMC hardware architecture for optimal hardware efficiency and accuracy.

\section{Analysis of Crossbar Size in IMC Hardware} \label{motivation}

\begin{figure*}[t]
\centering
\includegraphics[width=0.9\textwidth]{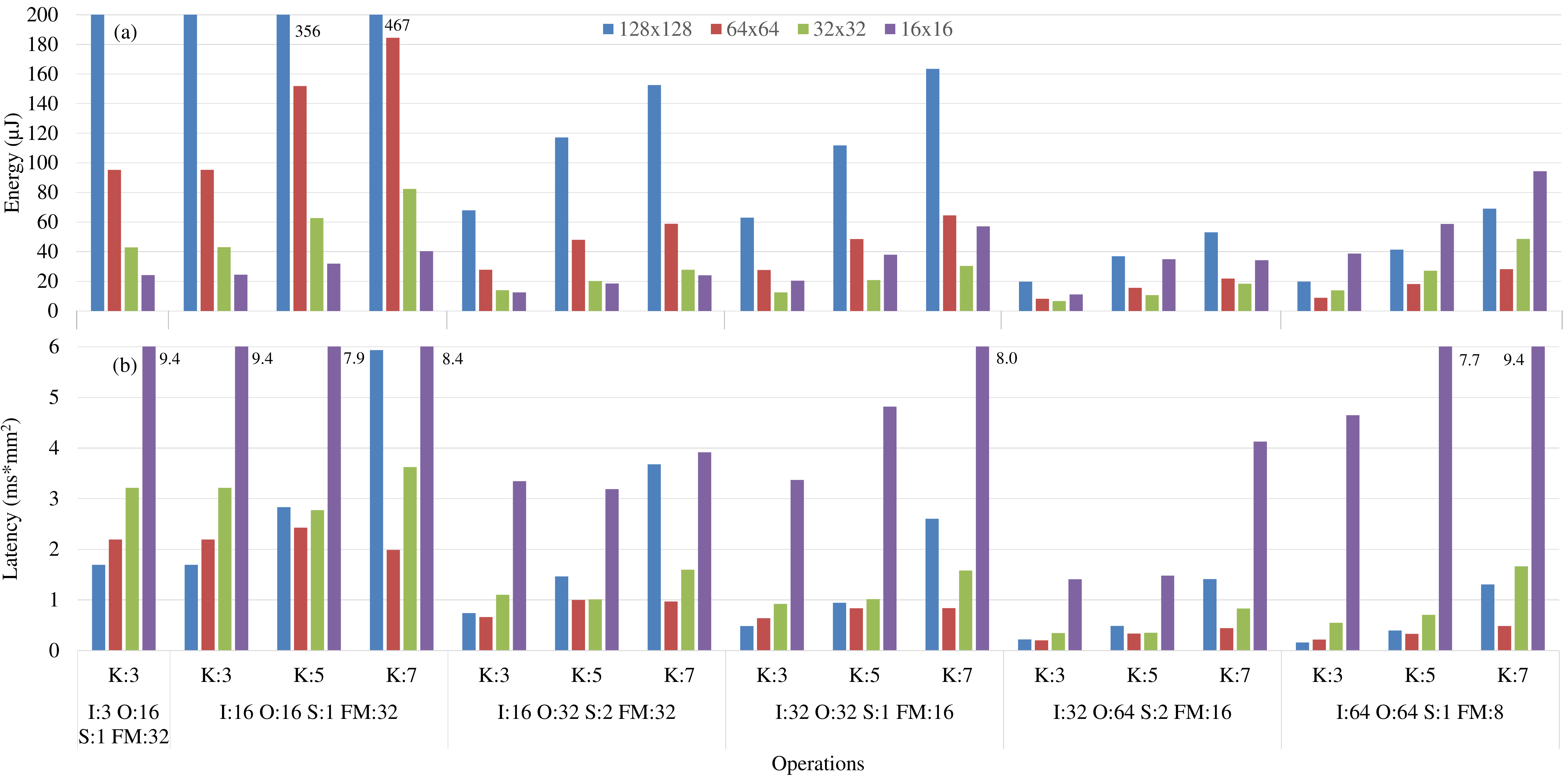}
\vspace*{-3mm}
\caption{Hardware Efficiency for different convolution operations mapped to varying crossbar sizes: (a) Inference energy (b) Inference latency (area normalized). FM is the input feature map size for the operation.}
\centering
\label{hwdeff}
\end{figure*}

We have alluded to the point briefly that crossbar size is an important parameter in analog IMC hardware because it has significant implications on the hardware efficiency and accuracy of the DNN, and these implications are dependent on DNN model parameters.
We perform extensive analysis to study the co-dependence of DNN model parameters and crossbar size.


\subsection{Implication of Crossbar Size on Hardware Efficiency}\label{motivation_a}


We study the implications of crossbar size on hardware efficiency of a typical spatial architecture, such as PUMA \cite{ankit2019puma}. 
PUMA's spatial architecture is organized hierarchically into three-tiers: cores, tiles and nodes. The basic computation unit inside the deepest hierarchy, that is the core, is a Matrix-Vector Multiplication Unit (MVMU). MVMU consists of a certain number of analog crossbars that are used to perform matrix vector multiplication operation. Multiple cores are connected via shared memory to constitute a tile. Finally, multiple tiles are connected via an on-chip network to constitute a node. 

A typical mapping of a DNN layer to PUMA's architecture is shown in Fig.~\ref{underutilization}. Let's say an input activation (IA) of dimension $H_{i} \times W_{i} \times I$ is convolved with a kernel (W) of dimension $K \times K \times I \times O$ to get an output activation (OA) of dimension $H_{o} \times W_{o} \times O$. First, the kernels are flattened and stored as columns of a 2D matrix and the input feature maps (FM) are flattened to get the input vectors to MVMUs. 
As the crossbar size is limited due to technology constraints, the 2D weight matrix is partitioned to map small chunks (of crossbar size) to MVMUs. This mapping results in some of the MVMUs being underutilized as illustrated in Fig.~\ref{underutilization}. Under-utilized crossbars still involve accesses to peripherals, despite not processing useful information. Thus under-utilization can lead to higher energy consumption and latency. The percentage of under-utilization in a crossbar depends on crossbar size and output channels. Hence, the hardware efficiency of different operations in a DNN depends on crossbar size they are mapped to as shown in Fig.~\ref{hwdeff}. 


We observe, in Fig.~\ref{hwdeff} (a) that the inference energy for different operations in a DNN have varying trends based on the crossbar size they are mapped to. For instance, the energy for operations where the number of output channels $O$ is low (say 16), decreases as we decrease the crossbar size from $128\times128$ to $32\times32$.
This occurs due to the under-utilization of the larger crossbars resulting in additional accesses to peripherals. As the number of output channels increases (to 32 or 64), the effect of under-utilization decreases and the smaller crossbars start to consume more energy compared to larger crossbars. This is because, when a DNN layer operation fully utilizes the crossbar, a smaller crossbar size leads to requirement of more MVMUs and consequently more peripheral accesses to represent the same operation.



The inference latency (area normalized) for the operations mapped to different crossbar sizes is shown in Fig.~\ref{hwdeff} (b). To understand the dependence of latency (area normalized) on crossbar size, we need to understand the dependence of area and latency in isolation. For example, like energy, latency has similar trends with crossbar size, i.e., for DNN operations with less output channels, lower crossbar sizes ($16\times16$) result in lower latency, whereas, for operations with more output channels, higher crossbar size leads to lower latency. However, area consumed by the compute core follows a different trend. In general, as crossbar size increases, number of required MVMUs reduce, resulting in reduced peripheral area overhead. But as the number of MVMUs per tile can be limited in a spatial architecture such as PUMA, for operations requiring a high number of MVMUs (K:7 I:16 O:16 S:1), the mapping may exceed the boundary of a tile, and require multiple tiles. In such a case, a larger crossbar ($128\times128$ here) size may end up requiring more area.

\subsection{Implication of Crossbar Size on Functionality of IMC Hardware}\label{motivation_b}

\begin{figure}[t]
\centering
\includegraphics[width=0.4\textwidth]{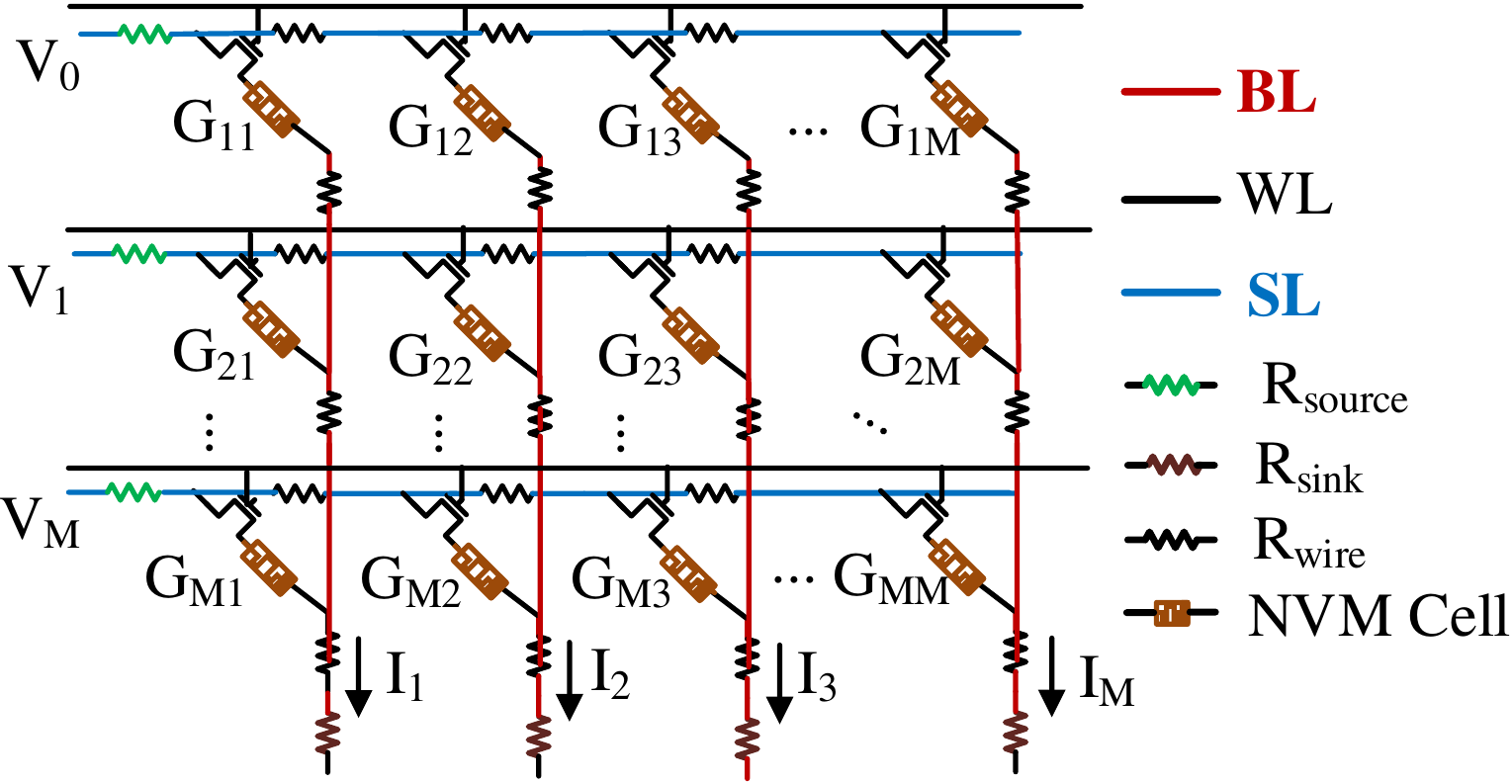}
\vspace*{-3mm}
\caption{Crossbar array structure with NVM cell at the crosspoints and non-ideal elements like parasitic resistance and access transistor.}
\centering
\label{xbarch}
\end{figure}

Next, we study the implications of crossbar size on the computational accuracy of IMC hardware. Fig.~\ref{xbarch} describes the basic structure of an MCA. There are several sources of non-idealities in an MCA, such as the peripheral resistances ($R_{source}$, $R_{sink}$), parasitic resistances ($R_{wire}$) and non-linear device I-V characteristics.
In order to perform an MVM operation in an MCA, the inputs are encoded as voltage (\textit{$V_{i}$}) and the weights are encoded as conductance of the device\textit{($G_{ij}$)}. 
The currents at the output of $j^{th}$ BL is $\textit{$I_{j}$} = \sum_{i}V_{i}G_{ij}$. 
However, due to the aforementioned non-idealities, the output current in an MCA is represented as:
\begin{equation}
    I = f(V, G(V), R_{source}, R_{sink}, R_{wire}) 
\end{equation}
where the device non-linear I-V characteristics is expressed such that the conductance ($G(V)$) is a function of input voltage ($V$). Hence to estimate the accuracy drop in a DNN due to these non-idealities we use GENIEx \cite{chakraborty2020geniex} that models the non-idealities of real crossbars using a neural network. 

\begin{figure}[t]
\centering
\includegraphics[width=0.4\textwidth]{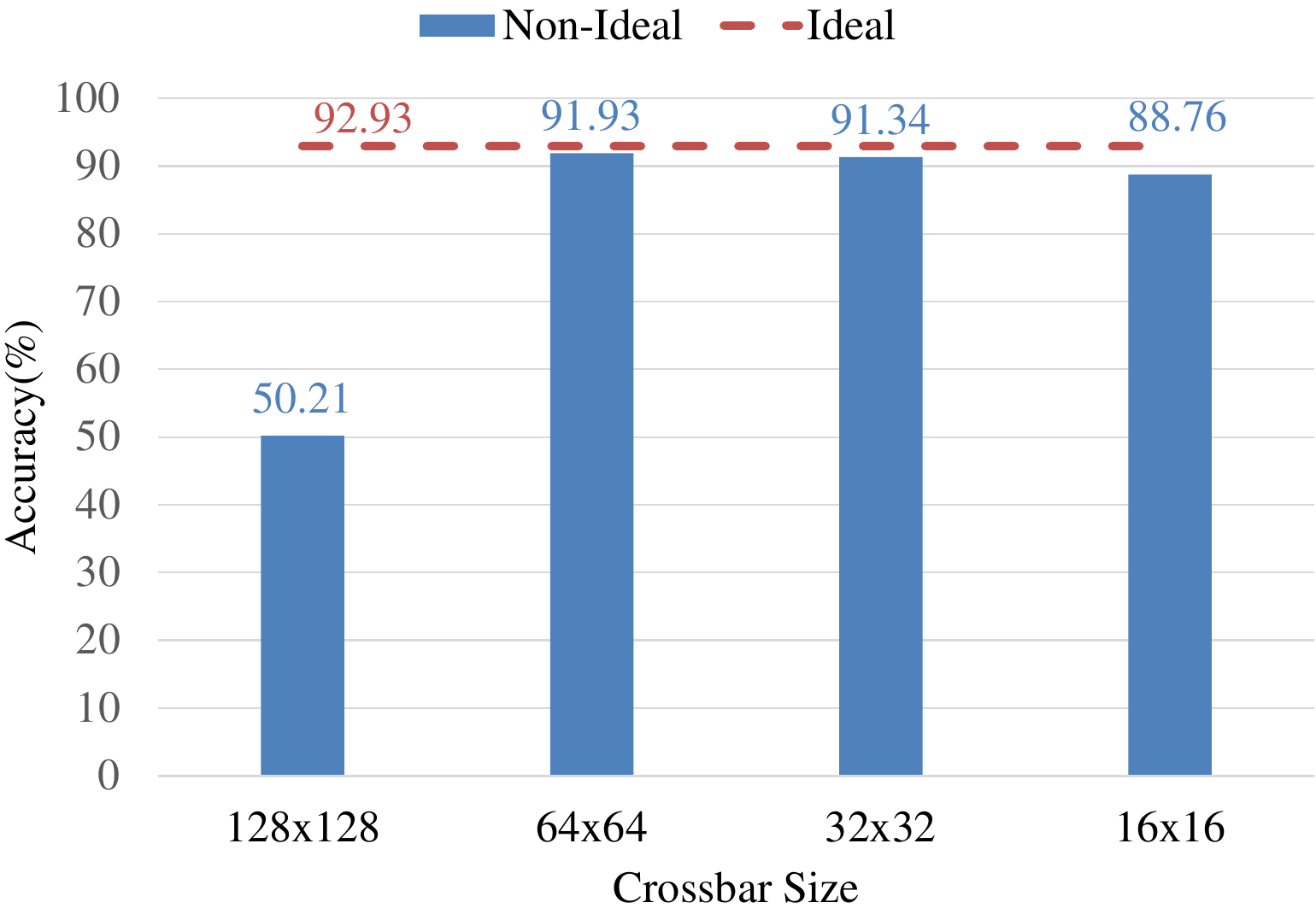}
\vspace*{-3mm}
\caption{Accuracy drop due to non-idealities in crossbar for ResNet-20 model mapped to different crossbars on CIFAR-10 dataset.}
\centering
\label{baselineacc}
\end{figure}

To understand the implication of crossbar size on the accuracy drop of a DNN due to the non-idealities in MCA, we map ResNet-20 model to different crossbar sizes based IMC hardware and the accuracy drop is shown in Fig.~\ref{baselineacc}. We make the following conclusions that can be made from this analysis: 

\begin{itemize}
    \item Large crossbars (128x128 here) incurs maximum drop in accuracy, due to longer metal lines resulting in higher $R_{wire}$ which leads to reduced effective resistance of the crossbar.
    \item As the crossbar size decreases the effect of non-idealities decreases, hence lower drop in accuracy. But when the crossbar size is lowered even more (16x16 here), the accuracy drop increases that occur due to the effect of non-linear non-idealities becoming more prominent compared to linear non-idealities. 
\end{itemize}

Hence, both the hardware efficiency and accuracy for DNNs mapped to the IMC hardware depend on crossbar size as well as layer operations. The unique trends of energy and latency (area normalized) with different operations motivate their co-optimization to get low energy-delay-area product (EDAP). To that effect, we propose \papername, a NAS-based approach to explore the search space presented by the unique co-dependence of DNN model parameters and crossbar size in IMC hardwares. Hereafter, area-normalized latency is referred to as latency.



\section{Proposed Approach}\label{section4}

\begin{figure*}[t]
\centering
\includegraphics[width=0.9\textwidth]{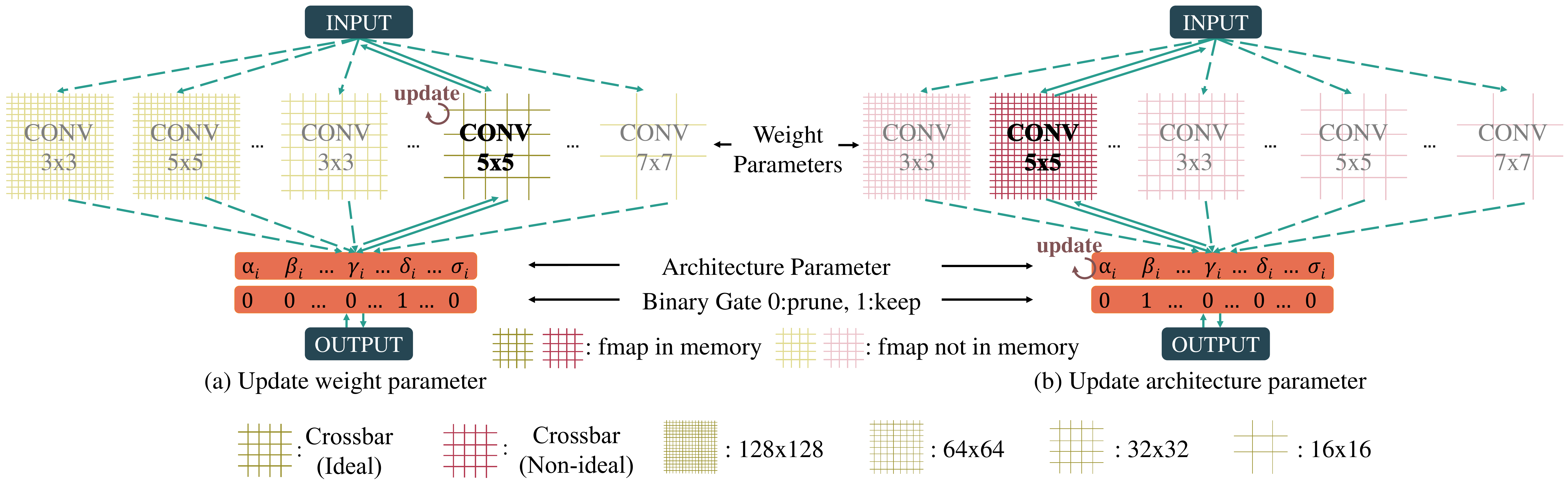}
\caption{An overview of \papername approach (best viewed in color): (a) First considering the crossbars without non-idealites, weights are trained keeping architecture parameters constant. (b) Next the architecture parameters are trained keeping the weight parameters constant and considering crossbars with non-idealities.}
\centering
\label{overparamnetwork}
\end{figure*}

In this section, we explain the methodology adopted by \papername. 
First, we introduce the search space for IMC hardwares and show how we consider the non-idealities in crossbar during the architecture search. Then, we show the latency and energy regularized loss function.

\subsection{Building Super-Network for NAS}\label{section41}

The proposed hardware-aware NAS framework, \papername, derives motivation from \cite{cai2018proxylessnas}. First, we start by describing an over-parameterized network to represent different types of operations.
The over-parameterized network is denoted by $\mathcal{N}(e_{1},...,e_{n})$ where $e_{i}$ represents a certain edge in the directed acyclic graph (DAG). Let $\mathcal{O} = \{o_{i}\}$ denote the candidate space for the over-parameterized network. The edge $e_{i}$ in the super-network is set as a mixed operation that has N parallel paths denoted as $m_{\mathcal{O}}$. Hence, the over-parameterized network can be expressed as $\mathcal{N}(e_{1}=m_{\mathcal{O}}^1,...,e_{n}=m_{\mathcal{O}}^n)$. 

As described in \cite{cai2018proxylessnas}, we also adopt path-level pruning which force only one path per layer to be active at run-time. This path-level pruning helps to reduce the memory requirement by an order and helps to search directly on large-scale datasets. If we look at path binarization in terms of equations, corresponding to the N parallel paths in a edge there are N real-valued architecture parameters (${\alpha_{i}}$) which are transformed to binary gates as shown in Eq.~(\ref{eq1}), where $p_{i} = \frac{exp(\alpha_{i})}{\sum_{j} exp(\alpha_{j})}$. For a given input $x$ the output of the mixed edge is defined as shown in Eq.~(\ref{eq2}). For more technical details, please refer to \cite{cai2018proxylessnas}.
\setlength{\abovedisplayskip}{3pt}
\setlength{\belowdisplayskip}{3pt}
\begin{equation}\label{eq1}
        g = \text{binarize}(p_{1}, ..., p_{N}) =
    \begin{cases}
        [1, 0, ..., 0],& \mbox{w/ probability $p_{1}$},\\
        ...\\
        [0, 0, ..., 1],& \mbox{w/ probability $p_{N}$}
    \end{cases}
\end{equation}

\begin{equation}\label{eq2}
    m_{\mathcal{O}}^{Binary}(x) = \sum_{i=1}^{N} g_{i}o_{i}(x) =
\begin{cases}
    o_{1}(x),& \mbox{w/ probability $p_{1}$},\\
    ...\\
    o_{N}(x),& \mbox{w/ probability $p_{N}$}
\end{cases}
\end{equation}


\subsection{Building and Training Over-Parameterized Network for IMC Hardware}\label{section42}
The over-parameterized network for an IMC hardware is unique due to the dependence of both accuracy and hardware efficiency on crossbar size as discussed in section \ref{motivation}. Hence, the candidate space $\mathcal{O}$ contains the operations mapped to different crossbar sizes and an extra \textit{zero} operator that helps to find the depth of the neural network. A particular edge of the over-parameterized network is shown in Fig.~\ref{overparamnetwork}. 

The training procedure of the over-parameterized network involves two steps first training the weight parameters and then the architecture parameters. To train the weight parameters the architecture parameters are frozen and one path from all the edges of $\mathcal{N}$ is stochastically sampled according to Eq.~(\ref{eq1}) and then the weight parameters are trained with standard gradient descent on the training set (Figure \ref{overparamnetwork} (a)). Next, to train the architecture parameters the weight parameters are frozen, then the binary gates are reset and the architecture parameters are updated on validation set (Fig.~\ref{overparamnetwork} (b)). To consider the effect of non-idealities in crossbar on accuracy, the inference through the selected architecture during the "update architecture parameter" step is done through GENIEx \cite{chakraborty2020geniex}. These two steps are performed alternately for certain number of epochs. Finally, to get the final compact neural network from $\mathcal{N}$, we select the path in each edge with the maximum architecture parameter value.

\subsection{Hardware Efficiency Regularized Loss Function}
Besides accuracy and latency, energy is an important parameter for spatial architectures. As shown in section \ref{motivation_a} both latency and energy of different operations mapped to IMC hardware vary with crossbar size and their trend for different operations are unique. Hence, when designing efficient neural network architecture for IMC hardwares, considering both latency and energy as objectives in the loss function is important. We follow the differentiable approach proposed in \cite{cai2018proxylessnas} to optimize both latency and energy. The expected hardware efficiency (energy or latency) of $i^{th}$ edge $e_{i}$ of $\mathcal{N}$ as a function of path weight $p_{j}^{i}$ is shown in Eq.~(\ref{eq3})

\begin{equation}\label{eq3}
    \mathbb{E}[HWE_{i}] = \sum_{j}p_{j}^{i} \times F(o_{j}^{i})
\end{equation}

\begin{figure*}[t]
\centering
\includegraphics[width=0.9\textwidth]{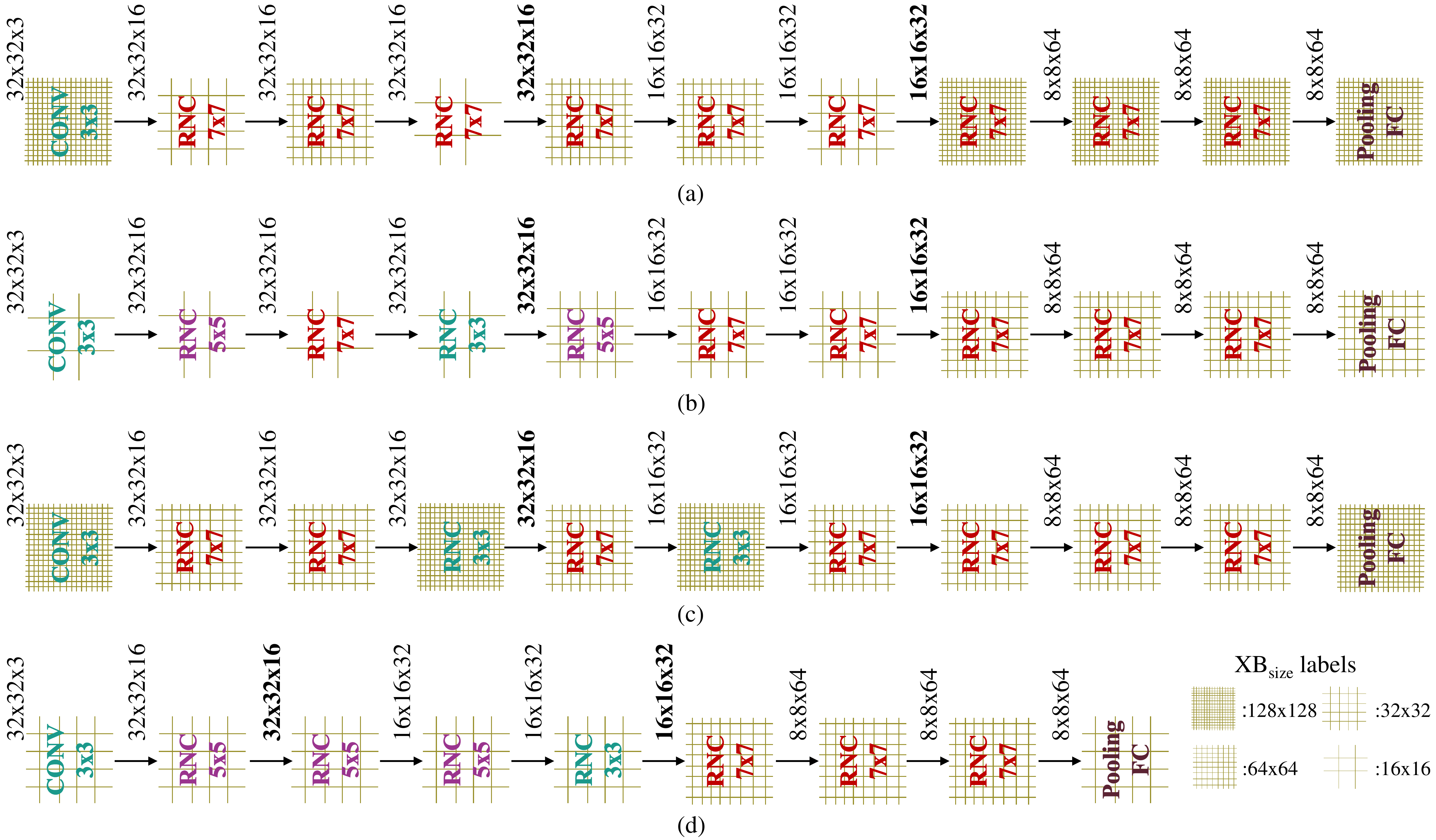}
\vspace*{-3mm}
\caption{Neural network and crossbar size found on CIFAR-10 with i-search when optimizing (a) None of hardware efficiency parameter (b) Energy (c) Latency (d) Energy and Latency. RNC represents the basic block of ResNet-20 model with two convolution layers in series and a skip connection. Feature map size in bold represents an input to the layer with stride = 2 block.}
\centering
\label{ablationarch}
\end{figure*}

where HWE can be latency or energy and F(.) denotes the hardware efficiency prediction model, which is a lookup table containing latency and energy for all the operations in the search space.  

Therefore, the expected hardware efficiency for $\mathcal{N}$ can be expressed by adding this expected hardware efficiency for all the layers of over-parameterized network:

\begin{equation}\label{eq4}
    \mathbb{E}[HWE] = \sum_{i}\mathbb{E}[HWE_{i}]
\end{equation}

Now we add the weighted latency and energy regularized terms to our loss function. The constants $\lambda_{1}$ and $\lambda_{2}$ controls the trade-off between accuracy, energy and latency. Hence the hardware efficiency regularized loss function can be expressed as:  
\begin{equation}\label{eq5}
Loss = Loss_{CE} + \lambda_{1}\norm{w}_{2}^2 + \lambda_{2}\mathbb{E}[latency] + \lambda_{3}\mathbb{E}[energy]
\end{equation}

where $Loss_{CE}$ is cross-entropy loss and $\lambda_{1}\norm{w}_{2}^2$ denotes weight decay.

\section{Experiments}\label{experiments}

We perform experiments on CIFAR-10 and Tiny ImageNet datasets to demonstrate the effectiveness of \papername to co-design neural network and hardware architecture. In addition, by an ablation study on CIFAR-10 dataset we show the importance of energy and latency co-optimization for MCA based IMC hardware. The IMC hardware parameters used in the experiments are shown in Table~\ref{table1}. All the experiments for \papername are conducted on four NVIDIA RTX 2080Ti GPUs.

\begin{table}[htb]
    \centering
    \caption{IMC Hardware Parameters, N is the number of rows in a crossbar.}
    \label{table1}
    \vskip 0in
    \begin{tabular}{|c|c|}
        \hline
        \textbf{Parameter} &  \textbf{Value} \\
        \hline\hline
         Bit stream &  1 bit\\
         \hline
         Bit slice & 2 bit\\
         \hline
         ADC precision & $\log_2 (N*3)$\\
         \hline
         $R_{on}$ & 100 k$\Omega$ \\
         \hline
         $R_{off}$ & 600 k$\Omega$ \\
         \hline
         $V_{supply}$ & 0.25 V \\
         \hline
    \end{tabular}
\end{table}

\subsection{Experiments on CIFAR-10}\label{experiments_a}
For the CIFAR-10 experiments, we start with ResNet-20 as our backbone. To create the over-parameterized network we replace the convolution layers of basic block in ResNet-20 model with the mixed layer which consists of different kernel sizes \{3x3, 5x5, 7x7\} mapped to varying crossbar sizes ($XB_{size}$) \{128x128, 64x64, 32x32, 16x16\}.
\begin{table}[h]
    \centering    
    \caption{Ablation study on CIFAR-10 with mono/bi/multi-objective loss function.}
    \label{table2}
    \vskip 0.0000001in
    \begin{tabular}{|c|c|c|c|c|}
        \hline
        \textbf{\thead{HWE Parameter\\ Optimized}} & \textbf{\thead{Energy\\ ($mJ$)}}  & \textbf{\thead{Latency\\ ($s*mm^2$)}} & \textbf{\thead{i-accuracy\\ ($\%$)}} \\
        
        \hline\hline    
         None & 1.59 & 1.27 & 93.18\\
         \hline
         Energy & \textcolor{red}{0.54}& 1.08 & 93.43\\
         \hline
         Latency  &  1.94 & \textcolor{red}{0.44} & 93.00\\
         \hline
         Latency \& Energy & \textcolor{red}{0.44} & \textcolor{red}{0.25} & 93.05\\
         \hline
    \end{tabular}
\end{table}


\textbf{Searching and training details:} From the training set, we create a validation set by randomly sampling 5000 images. This subset is used to learn the architecture parameters using Adam optimizer with an initial learning rate of 0.006. We consider two search configuration in \papernamewospace: i) non-ideal search (\textit{ni-search}), where the inference while training the architecture parameters is performed considering non-idealities in the crossbars, ii) ideal search (\textit{i-search}), where we do not consider non-idealities in crossbars. The functional simulator in GENIEx uses bit-serial compute units to implement the MVM operation which is further used to create custom \textit{conv2d} and \textit{linear} layers operation libraries. Hence, inference during ni-search has high memory and time requirements. To address the high time requirement we provide tiling support for the output activations and multi-GPU support for the whole GENIEx framework which gives $\sim$5-6x speedup compared to the original version. The added tiling feature also helps to create a trade-off between batch size for inference (while training architecture parameters) and the tile size. Even with all the above speedups, the time for inference during ni-search is high hence at every iteration we randomly select 2 layers from the over-parameterized network mapped to crossbars with non-idealities; the first convolution layer and last fully-connected layer are always mapped to crossbars with non-idealities.

\begin{table*}[!htb]
    \footnotesize
    \centering    
    \caption{Accuracy and hardware efficiency on CIFAR-10. $XB_{size}$ = mixed implies varying crossbar size across the layers.}
    \label{table3}
    \vskip 0.0000001in
    \begin{tabular}{|c|c|c|c|c|c|c|}
        \hline
        \textbf{\thead{Model\\ ($XB_{size}$)}} &  \textbf{\thead{i-accuracy\\ ($\%$)}} & \textbf{\thead{ni-accuracy\\ ($\%$)}} & \textbf{\thead{Energy\\ ($mJ$)}} & \textbf{\thead{Latency\\ ($s*mm^2$)}} & \textbf{\thead{EDAP*1e3\\ ($mJ*ms*mm^2$)}} & \textbf{\thead{Params\\ ($M$)}} \\
        \hline\hline    
    
         ResNet-20 (128x128) & 92.93 & 50.21 & 2.04 & 0.32 & 0.65 & 0.27\\
         \hline
         ResNet-20 (64x64) & 92.93 & 91.93 & 0.89 & 0.40 & 0.35 & 0.27\\
         \hline
         ResNet-20 (32x32) & 92.93 & 91.34 & 0.45 & 0.61 & 0.27 & 0.27\\
         \hline   
         ResNet-20 (16x16) & 92.93 & 88.76 & 0.49 & 2.14 & 1.05 & 0.27\\
         \hline
         \papernamewospace-i (mixed) & 93.05 & 92.13 & 0.44 & 0.25 & 0.11 & 1.23\\
         \hline
         \papernamewospace-ni1 (mixed) & 93.39 & 92.92 & 0.82 & 0.40 & 0.33 & 1.23 \\
         \hline
         \papernamewospace-ni2 (mixed) & 93.15 & 92.70 & 0.78 & 0.38 & 0.29 & 1.22 \\
         \hline
         \papernamewospace-ni3 (mixed) & 93.45 & 92.22 & 0.51 & 0.38 & 0.19 & 0.94 \\
         \hline
    \end{tabular}
\end{table*}

\begin{figure*}[!htb]
\centering
\includegraphics[width=0.9\textwidth]{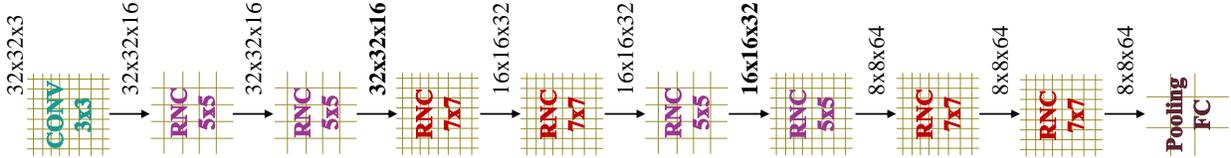}
\vspace*{-3mm}
\caption{Neural network (\papernamewospace-ni2) and crossbar size (label same as Fig.~\ref{ablationarch}) found on CIFAR-10 with ni-search when optimizing both energy and latency. RNC represents the basic block of ResNet-20 model with two convolution layers in series and a skip connection. Feature map size in bold represents an input to the layer with stride = 2 block.}
\centering
\label{arch}
\end{figure*}

After the training process of the over-parameterized network is completed we select a compact network according to the architecture parameter values as discussed in section~\ref{section42}. Then we train this compact network with step learning rate schedule for 300 epochs with learning rate starting from 0.1 and divide by 10 at 100th and 150th epoch. We report the accuracy on test set considering non-idealities in crossbar (\textit{ni-accuracy}) and without considering non-idealities in crossbar (\textit{i-accuracy}).

\textbf{Ablation Study:} To motivate the co-optimization of energy and latency, we find neural network and hardware architecture configuration using i-search in \papername by optimizing: i) none of the hardware efficiency (HWE) parameter (none), ii) one of the hardware efficiency parameter (energy or latency), iii) both energy and latency. The result for the ablation study is tabulated in Table~\ref{table2} and the findings are summarized below:

\begin{itemize}
    \item When none of the hardware efficiency parameter is optimized, energy and latency values are high; the neural network and crossbar size found is shown in Fig.~\ref{ablationarch} (a).
    \item When energy/latency is considered in the loss function, only those parameters are optimized; the neural network and crossbar size found is shown in Fig.~\ref{ablationarch} (b)/(c) respectively and the crossbar size for different layers are selected to optimize the respective hardware efficiency parameters.  
    \item Finally, when both energy and latency are considered in the loss function, all the hardware efficiency parameters are optimized. To optimize all the hardware efficiency parameters together for some of the layers the NAS algorithm selects \textit{zero} operation from the search space which is depicted in Fig.~\ref{ablationarch} (d). 
\end{itemize}

\begin{table*}[!htb]
    \footnotesize
    \centering    
    \caption{Accuracy and hardware efficiency on Tiny ImageNet. $XB_{size}$ = mixed implies varying crossbar size across the layers.}
    \label{table4}
    \vskip 0.0000001in
    \begin{tabular}{|c|c|c|c|c|c|c|c|}
        \hline
         \textbf{Model ($XB_{size}$)} & \multicolumn{2}{|c|}{\textbf{i-accuracy ($\%$)}} & \multicolumn{2}{|c|}{\textbf{ni-accuracy ($\%$)}} & \textbf{Energy ($mJ$)} & 
         \textbf{Latency ($s*mm^2$)} & 
         \textbf{EDAP*1e3 ($mJ*ms*mm^2$)} \\
        \cline{2-5}
        &\textbf{top-1} & \textbf{top-5} & \textbf{top-1} & \textbf{top-5}& & &\\
        \hline\hline    
    
         ResNet-18 (128x128) & 55.83 & 78.48& 34.35 & 59.64 & 2.31 & 0.83 & 1.91 \\
         \hline
         ResNet-18 (64x64) & 55.83 & 78.48 & 53.65 & 76.62& 1.97 & 1.37 & 2.71\\
         \hline
         \papernamewospace-i (mixed) & 56.57 & 78.70 & 44.76 & 69.20 & 1.99 & 0.85 & 1.71\\
         \hline
         \papernamewospace-ni1 (mixed) & 56.58 & 78.60 & 52.92 & 76.77 & 2.05 & 1.03 & 2.11\\
         \hline
         \papernamewospace-ni2 (mixed) & 56.32 & 78.83 & 53.81 & 77.68 & 2.31 & 1.11 & 2.59\\
         \hline
    \end{tabular}
\end{table*}

\begin{figure*}[!htb]
\centering
\includegraphics[width=0.9\textwidth]{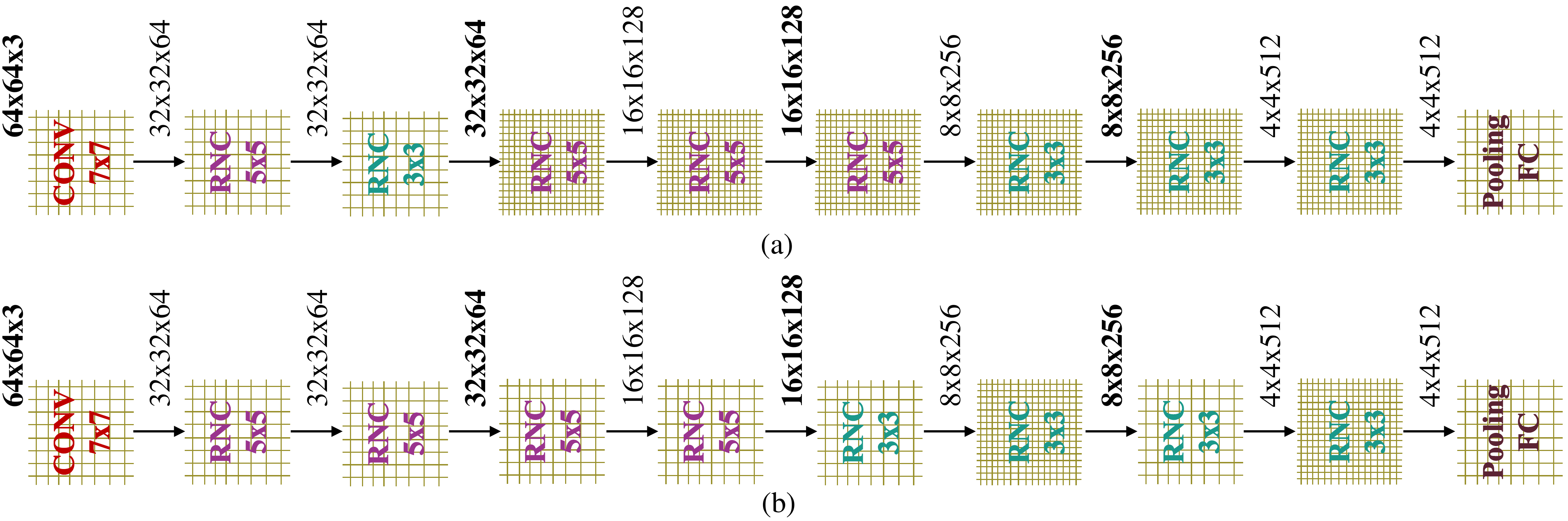}
\vspace*{-3mm}
\caption{Neural network and crossbar size (label same as Fig.~\ref{ablationarch}) found on Tiny ImageNet with (a) i-search (\papernamewospace-i) (b) ni-search (\papernamewospace-ni2) when optimizing both energy and latency. RNC represents the basic block of ResNet-18 model with two convolution layers in series and a skip connection. Feature map size in bold represents an input to the layer with stride = 2 block.}
\centering
\label{arch_imagenet}
\end{figure*}

\textbf{Non-ideal search results:} We apply the proposed approach in section~\ref{section42} to  find neural networks and crossbar size for different layers of the neural network. We compare the test accuracy and hardware efficiency results for the neural network architecture (\papernamewospace-ni) found by ni-search in \papername with the neural network (\papernamewospace-i) found by i-search in \papername and the baseline ResNet-20 model mapped to homogeneous crossbar sizes across all the layers. The results are tabulated in Table~\ref{table3} and the neural network and crossbar size found using i-search and ni-search is shown in Fig.~\ref{ablationarch} (d) and Fig.~\ref{arch} respectively. Compared to all the ResNet-20 models, both \textit{\papernamewospace-ni} and \textit{\papernamewospace-i} have better i-accuracy and ni-accuracy. The ni-accuracy for \papernamewospace-ni1-3 is better compared to \papernamewospace-i, which can be attributed to the GENIEx inference during the ni-search that helps to select operations and crossbar size from the candidate set having better ni-accuracy. The EDAP and accuracy for \papernamewospace-ni2 is 17\% lower and 0.8\% higher than the best ni-accuracy ResNet-20 model (ResNet-20 (64x64)). Further, the EDAP and accuracy for \papernamewospace-ni3 is 29.63\% lower and 0.9\% higher respectively than the best EDAP ResNet-20 model (ResNet-20 (32x32)). Also, both \papernamewospace-i and \papernamewospace-ni use more parameters compared to ResNet-20 model; however, the hardware efficiencies for the model found by \papername is much better compared to the baseline ResNet-20 models, which is due to the high utilization of crossbars in the architectures found by \papernamewospace. Moreover, this motivates optimizing hardware efficiency directly instead of proxy metric like number of parameters in the neural network. 

The ni-accuracy difference between \papernamewospace-ni and \papernamewospace-i models is not high, because the crossbar size in the CIFAR-10 search space that have high ni-accuracy (64x64, 32x32) also results in high hardware efficiency. These hardware efficiency and ni-accuracy results demonstrate the importance of exploring the search space proposed in section~\ref{section42} to co-design neural network and hardware architecture for MCA based IMC hardware. The search cost for ni-search is $\sim$32 GPU-hours and for i-search it is $\sim$12 GPU-hours.

\subsection{Experiments on Tiny ImageNet}\label{experiments_b}

For the Tiny ImageNet experiments, we start with ResNet-18 (max-pool layer after first convolution layer removed) \cite{he2016deep} as our backbone. Similar to CIFAR-10 experiments we build the super-network by replacing the convolution layers of basic blocks in ResNet-18 with the mixed layer that consists of different kernel sizes \{3x3, 5x5\} mapped to varying crossbar sizes \{128x128, 64x64\}. 

\textbf{Searching and training details:} We randomly sample 5000 images from the training set as a validation set to train the architecture parameters using Adam optimizer with an initial learning rate of 0.003. Similar to CIFAR-10 experiments we consider two search configurations ni-search and i-search. While training the architecture parameters, the stride=2 blocks, first convolution layer and last fully-connected layer in the over-parameterized network are always mapped to crossbars with non-idealities and additionally we randomly map 4 layers from rest of the layers in the over-parameterized network to crossbars with non-idealities at every "update architecture parameter" iteration. The compact network from the over-parameterized network is selected similar to CIFAR-10 experiments. We train this compact network with step learning rate schedule for 90 epochs with learning rate starting from 0.1 and divide by 10 at 30th and 60th epoch.

\textbf{Non-ideal search results:} We apply the \papername approach proposed in section~\ref{section42} to find neural network and crossbar size on Tiny ImageNet dataset. The results are listed in Table~\ref{table4} and the neural networks and hardware architecture configuration found using i-search and ni-search are shown in Fig.~\ref{arch_imagenet}. Compared to the best ni-accuracy ResNet-18 model (ResNet-18 (64x64)) \papernamewospace-ni2 has 0.2\% better accuracy with 4.4\% lower EDAP. Further, \papernamewospace-ni1 has 18.57\% higher accuracy with 10.47\% higher EDAP compared to best EDAP ResNet-18 model (ResNet-18 (128x128)).  

Majority of the layers in Fig. ~\ref{arch_imagenet}(a) use 128x128 crossbars (hence high hardware efficiency) because larger crossbars are fully utilized due to the output channels more than 128 for the operations in the Tiny ImageNet search space. Unlike CIFAR-10 the ni-accuracy difference between \papernamewospace-ni1,2 and \papernamewospace-i models is high because the operations in the Tiny ImageNet search space that have high ni-accuracy results in low hardware efficiency. This shows the importance of inference through GENIEx during the "update architecture parameter" step to find the crossbar size with better ni-accuracy. The search cost for ni-search is $\sim$350 GPU-hours and for i-search it is $\sim$5 GPU-hours. 

\section{Conclusion}
The efficiency of DNNs are co-dependent on DNN model parameters and the underlying hardware architecture. This is particularly true for memristive crossbar based in-memory computing systems. On one hand, these systems suffer from the effect of device-circuit non-idealities which govern the accuracy of DNNs, and on the other hand, the hardware efficiency is dominated by the peripheral ADC cost. Incidentally, the impact of non-idealities on DNN accuracy as well as the implications of ADC cost on hardware efficiency, is a strong function of the size of crossbars in IMC hardware. In this work, we present \papernamewospace, a NAS framework to co-design neural network and the underlying hardware architecture through co-exploration of DNN model parameters and crossbar-size for optimal hardware efficiency and accuracy. First, we analyze the implications of crossbar size on hardware efficiency and accuracy (considering non-idealities in crossbars) for different layers of the neural network comprising of different kernel sizes mapped to varying crossbar sizes. Further, we demonstrate the importance of co-optimization of accuracy, energy and latency to build neural networks with high accuracy and low EDAP for IMC hardware. Finally, we show that \papername can co-design neural network and IMC hardware to achieve 17\%, 4\% lower EDAP and 0.8\%, 0.2\% higher accuracy, respectively, on CIFAR-10 and Tiny ImageNet datasets with respect to baseline ResNet-20 and ResNet-18 networks. 

\section*{Acknowledgement}
This work was supported in part by the Center for Brain-inspired Computing Enabling Autonomous Intelligence (C-BRIC), one of six centers in JUMP, a Semiconductor Research Corporation (SRC) program sponsored by DARPA, in part by the National Science Foundation, in part by Intel, and in part by the Vannevar Bush Faculty Fellowship.



\bibliographystyle{IEEEtran}
\bibliography{conference_101719}

\end{document}